# Four-Point, 2D, Free-Ranging, IMSPE-Optimal, Twin-Point Designs

Selden Crary[1] and Jan Stormann[2]

*1. Palo Alto, USA     2. Lycée Français de Vienne, Austria*

**Abstract**

We report the discovery of a set of four-point, two-factor, free-ranging, putatively IMSPE-optimal designs with a pair of twin points, in the statistical design of computer experiments, under Gaussian-process, fixed-Gaussian-covariance-parameter, and zero-nugget assumptions. We conjecture this is the set of free-ranging, twin-point designs with the smallest number of degrees of freedom.

*In any field, find the strangest thing and then explore it.*
*-- John Archibald Wheeler*

## 1. History

### 1a. Overview

Our interest is statistical design for computer experiments, under the following, admittedly long, list of assumptions: fixed, natural number of points; fixed number of independent variables; Gaussian-process model; fixed Gaussian-covariance parameters; no nugget; and the IMSPE-optimal design criterion, which was expressed concisely, as the synonymous IMSE criterion, in Eq. 2.9 of Sacks, Schiller, and Welch (SSW) [1], as summarized in the Appendix.

Statistical practitioners, even when dealing with cases for which the IMSPE criterion would be well suited, often cite the following problems with IMSPE-optimal designs: the covariance matrix of SSW's Eq. 2.9 is ill-conditioned when any two points are proximal or there are a large number of points; searching for the optimal design is cpu-intensive; the resulting designs often have poor projective properties to lower numbers of independent variables, when some independent variables are irrelevant; the designs found often have clusters of points, which it is mistakenly and sometimes ineradicably thought, are of questionable value for prediction, as replicated points add no information to the problem and thus the second of two proximal points must be close to irrelevant [Ref. 2, pp. 2981-2982]; and gaps appear in regions of the designs, causing, it is again mistakenly thought, unavailability of information for prediction in these regions. Many dissertations and hundreds of papers have addressed these concerns and offered alternatives. We refer the interested reader to the following select list: dissertations [3-4]; recent, review paper [5].

*Digression:* We do not explicitly address the above concerns of practitioners, except for the following two corrections to the otherwise excellent review paper mentioned, above [5]. On p.

701, ¶ 2, it seems the assumption is made that proximal points cannot arise in the presence of an effective inter-point repulsion. This is false, because if repulsion grows with inter-point distance, in more than one independent variable, then two proximal points can be pushed together by the combined action of their distant relatives – a forced marriage, if you will. This effect was observed in the design with dot diagram given in Fig. 1, of the next sub-section below. The other correction is what seems to be a failure to recognize, on p. 690, Col. 2, ¶ 2, that the integral, for the important case of fixed Gaussian-covariance parameters, can be written in closed form in terms of error functions, thus obviating the need for approximating the integral via a discrete sum over a finite grid. We note that both Maple and Mathematica provide error-function evaluations to more than one-million digits.
*End of digression.*

### 1b. Free-ranging, IMSPE-optimal designs

Our group's strong interest in IMSPE-optimal designs began with the discovery of a pair of highly proximal points dubbed "twin points" [6-8], in the 2D, N=11, IMSPE-optimal design problem on the design domain $[-1,1]^2$, with covariance parameters $[\theta_1,\theta_2]=[0.128,0.069]$. The design, which had IMSPE=0.00000502762…, is shown in Fig.1, below.

| $x_1$ | $x_2$ |
|---|---|
| -0.8590… | 0.0014… |
| -0.7921… | 0.7796… |
| -0.7826… | -0.8038… |
| -0.3144… | 0.0395… |
| 0.0000… | 0.9069… |
| 0.3144… | 0.0395… |
| 0.7826… | -0.8038… |
| 0.7921… | 0.7796… |
| 0.8590… | 0.0014… |
| -0.0000… | -0.6171…+δ |
| -0.0000… | -0.6171…-δ |

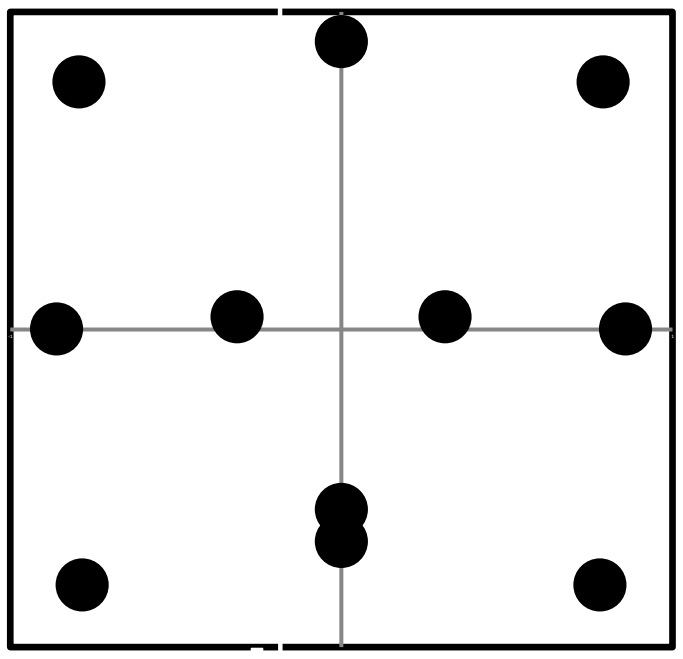

Fig. 1. Design listing (left) and dot diagram (right), plotted with abscissa $x_1$ and ordinate $x_2$, and with exaggerated twin separation, are shown for the first, putatively optimal, twin-point design.

For the design in Fig. 1, or its sibling formed by mirror reflection about the abscissa, the IMSPE has the following properties: a parabolic minimum, local to the twins, as the distance between the twins is allowed to decrease along the $x_2$ axis; a parabolic maximum, local to the twins, as the distance between the twins is allowed to decrease along the $x_1$ axis; and $C^\infty$ continuity everywhere, except at an essential discontinuity in the zero-separation limits. In very close proximity, there are two cases to consider, depending upon whether directional-derivative information is available explicitly, via adjoint or other methods [3]. Each case represents a distinct homotopy class. Fig. 2, below, is a 3D plot of the IMSPE, its essential discontinuity, and the jump in IMSPE, when directional-derivative information is not available.

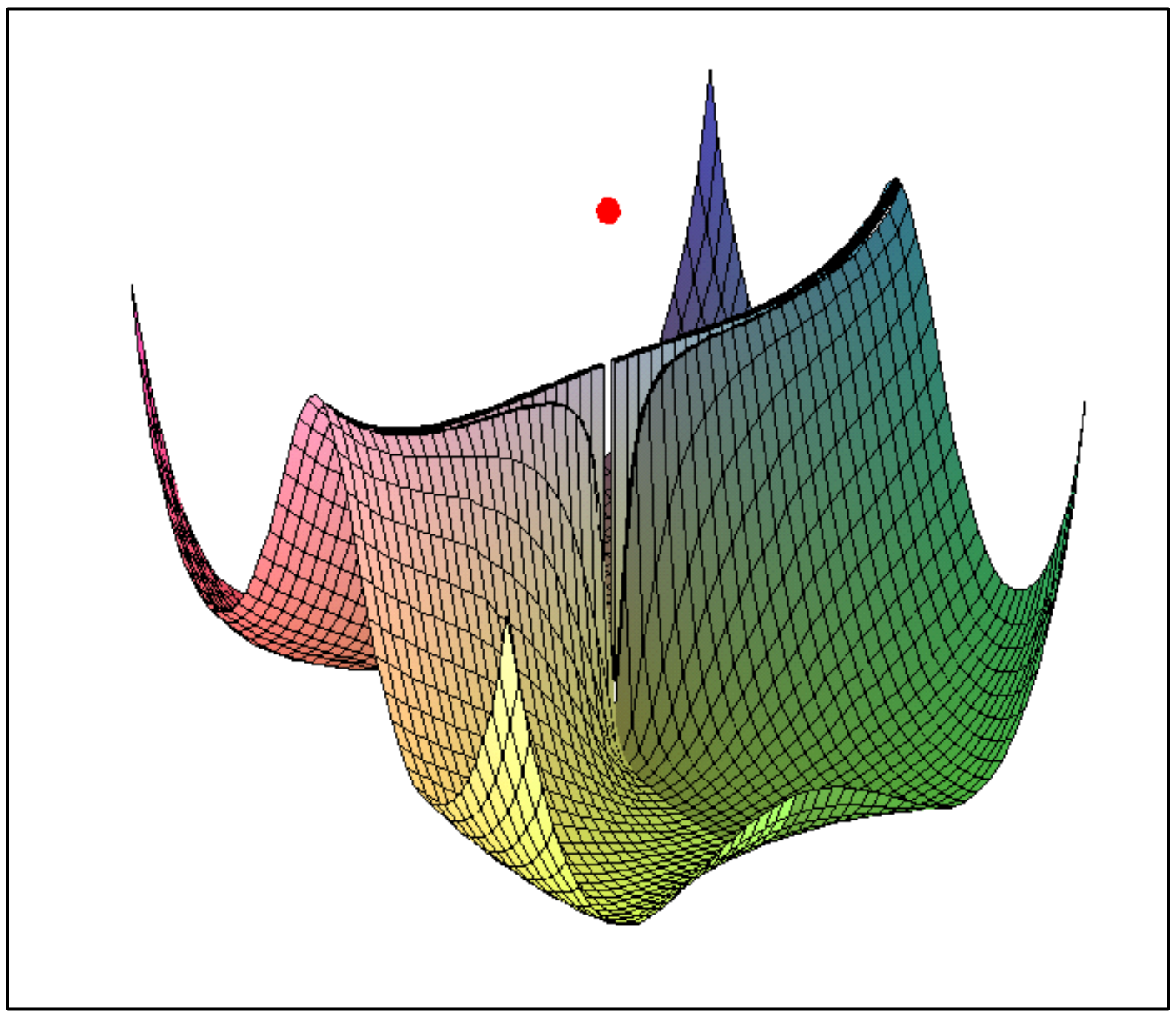

Fig. 2. This 3D plot shows the IMSPE of the design in Fig. 1, plotted vertically, as a function of the vector position of one of its twin points, along an obvious base, with the twins' midpoint and all other points fixed. The essential features are the following: (a) a minimum, in the zero-separation limit, and along a low-lying rift valley, when the twins are aligned along the $x_2$ axis; (b) a local maximum, in the zero-separation limit, and along a high-lying ridge, when the twins are aligned parallel with the $x_1$ axis; and (c) continuity of IMSPE and all its derivatives, everywhere, except for an essential discontinuity, along a continuous vertical locus from the minimum to the local maximum mentioned in (b). What appears as a cut in the 3D plot is an artifact of the plotting software and should be disregarded. The twins can merge in either of two ways, viz. without or with directional-derivative information available. The red dot represents the value of the IMSPE, in the former case, after the merge, in which case the IMSPE is that of the resultant N=10 design. In the other case, after the merge, there is no discontinuous upward jump, but the essential discontinuity still exists.

The presence of twin-points was unexpected, even "strange," to use a word in the famous John Wheeler quote, above, but under scrutiny it was clear that the points should be as considered as the following instruction: (a) If explicit directional-derivative information is available, then use both a function evaluation and the appropriate directional derivative, evaluated at the twin point, or (b) if directional-derivative information is not available, then use both the average of the responses and the appropriate directional difference, evaluated at the twins' midpoint.

Our group observed that clumping is common in IMSPE-optimal designs.

Subsequent theoretical work showed that such designs could arise, despite serious ill-conditioning of the covariance matrix used in an intermediary step of the conventional formula for IMSPE, through perfect cancellation of the singular terms, once the full IMSPE formula was completed in search software, via either hybrid symbolic-numerical computation or via sufficiently high-precision numerical computation [9].

### 1c. Free-ranging, clustered, MMSE-optimal sequential design

The interpretation of an instruction to use directional-derivative information was not new. In the context of optimal sequential designs, Sacks, Welch, Mitchell, and Wynn (SWMW) mentioned the concept, and they observed a "tendency for design sites to eventually 'pile up,' in a manner that 'may seem counter-intuitive'" [Ref. 10, p. 419, ¶ 3]. They then gave a non-free-ranging 1D, N=2 example, for the MMSE criterion, in which there was such piling up, but the authors were quick to point out that the design was not optimal in the non-sequential case. Our interpretation was that SWMW, too, found the proximal points "strange."

### 1d. Other examples of free-ranging clustering in optimal designs

The literature contains other reports of clustering, but none of these included investigations of the limits as the inter-point distance of proximal points went to zero from various directions. Optimal-design searches for variogram or covariance-parameter estimation showed expected clustering [11-14]. Given current thinking, it would be interesting to investigate the nature of these clusters to see what specific directional-derivative or directional-difference information they specified. Zhu and Stein observed clumps of points in designs for prediction [15]. Leatherman's recent dissertation included the following named design tables that showed clusters of points that, if investigated with higher-precision computation, may include not only twin points but triplet points and quadruplet points, as well: B.83 (possibly one set of twins); B.89 (possibly three sets of twins); and B.90 (possibly two sets of twins, three sets of triplets, and one set of quadruplets) [4].

### 1e. Overnight discovery of a non-free-ranging, twin-point system

In a Spring Research Conference 2012 presentation, the name "ε-clustered designs" was proposed for the designs, when no explicit directional-derivative information was present, and a moderately complex phase diagram for an N=11, two-factor system was demonstrated [16], along with additional theoretical work. At the following-day's Conference breakfast, Hickernell demonstrated, via a hybrid symbolic-numerical computation using Mathematica, his overnight discovery of a phase transition in a simple N=2, 1D, IMSPE-optimal problem, from finitely spaced points above a critical value of θ, to a twin-point design below the critical θ, albeit with one of the points held fixed at the center of the design domain [16-17].

## 2. Attitude

For the research reported here, we feigned disinterest both in statistical practice and the expense of high-precision cpu-intensive computation.

## 3. Motivation

Due to the unavailability of adequately high-precision software for evaluation of both the error function arising in IMSPE evaluations, as well as for the needed downhill-search algorithm, there has not been any report, independent of the authors' group, of free-ranging, ε-clustered designs. This fact leads to this paper's two motivations: to provide a convincing demonstration that free-ranging, IMSPE-optimal, clustered-point designs exist and to provide software with which others can make their own observations leading to the same conclusion.

## 4. Outline

Section 5 announces the availability of software with the needed precision; Section 6 announces our discovery of a twin-point, N=4 design in two factors; Section 7 is a summary of this paper's key results; Section 8 provides conjectures for future research; Section 9 includes some concluding comments; and Section 10 is a revision history.

## 5. Software

Based on algebra detailed in the Appendix, we developed the following, easily transportable Maple programs, which allow sufficiently high precision for evaluating designs and for searching for IMSPE-optimal designs, including those with ε-clustered points, over the domain $[-1,1]^2$:

*evalIMSPE2D* evaluates a design's IMSPE.

*minIMSPEccd2D* uses a cyclic-coordinate-descent algorithm to find local or global IMSPE minima.

Any responsible party interested in obtaining a copy of any Maple program or Microsoft-Excel spreadsheet used in the generation of any of the figures in this paper, or of any of the figures themselves, is welcome to contact the first author, at email address: selden_crary at yahoo dot com.

## 6. N=4, two-factor, free-ranging optimal designs

We used *minIMSPEccd2D* to search for N=4, free-ranging, optimal designs over a wide range of covariance-parameter pairs and constructed the IMSPE-vs.-$\theta_2$ phase diagram shown in Fig. 3, below, where thin solid lines are constant-$\theta_1$ parametric curves, and thick solid (resp., dashed) lines represent discontinuous (resp., continuous) jumps of the design points as $\theta_2$ is varied and phase boundaries, between adjacent phases with distinct symmetries, are crossed. In particular, the dashed solid-red line represents the $\theta_1=\theta_2$ locus. Moving from left to right, in Fig. 3, and at constant IMSPE=0.001, the designs in each phase are characterized, as follows: 4-in-line, rhomboid with twin points, rhomboid, rectangle, square, rectangle, rhomboid, rhomboid with twin points, and 4-in-line. Details are given in the figure caption, and a representative design for each phase is shown in Fig. 4, below.

For covariance parameters $\theta_1=0.128$ and $\theta_2=0.00016$ we found the rhombus-with-twins design shown in Fig. 5, below, which had an IMSPE of 0.0000668211…. To establish confidence in the design's optimality, we evaluated the IMSPE of more than one-million, uniformly-random-generated N=4 designs with identical design domains and covariance parameters. Each design had an IMSPE greater than the IMSPE of the design of Fig. 5, thus providing strong evidence for the optimality of the design in Fig. 5. It also provided strong evidence, albeit short of a mathematical proof, that designs with proximal points can be IMSPE-optimal.

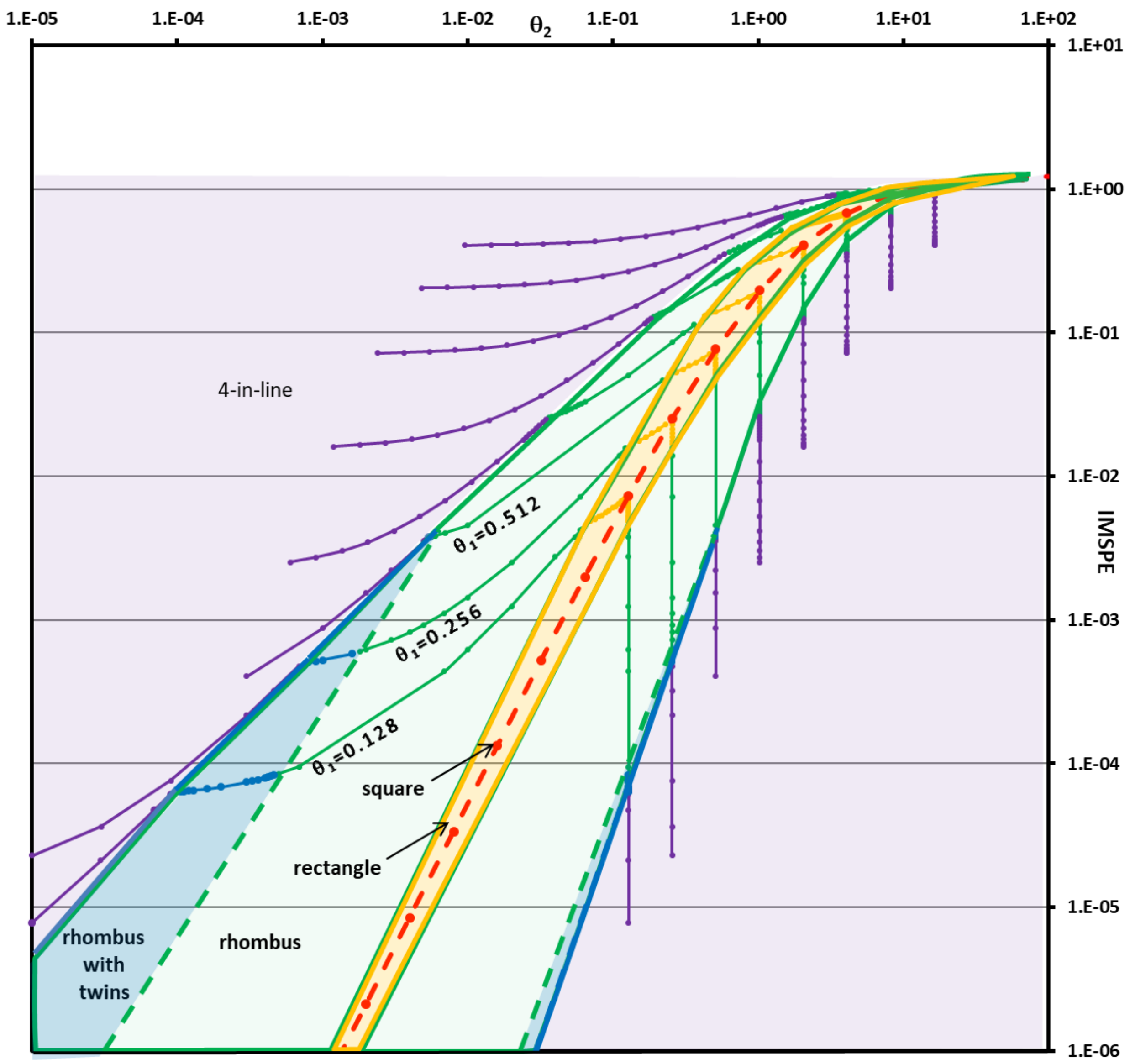

Fig. 3. The IMSPE-vs.-$\theta_2$ phase diagram for free-ranging, N=4, two-factor, IMSPE-optimal designs, with constant-$\theta_1$ parametric lines, is shown. Color key for phases on or to the left (resp., right) of the dashed red line: [dashed red line] centered axis-oriented squares; [yellow] centered rectangles, with bottoms aligned with the $x_1$ (resp., $x_2$) axis; [green] centered rhomboids, with top and bottom points on the $x_2$ (resp., $x_1$) axis; [blue] rhomboids, as above, but with twins on the $x_2$ (resp., $x_1$) axis; [purple] 4-in-lines, i.e. designs with all points on the $x_1$ (resp., $x_2$) axis. Small dots represent search results.

The parabolic variation of the IMSPE of the design of Fig. 5, as a function of twin separation, along each cardinal axes, is shown in Fig. 6, below. Although the focus of this paper is not on statistical practice, it may be of interest to practitioners that the IMSPE is nearly invariant to vertical twin separation. This means that the four design points may lie on a rhombus of any height, and with a center entirely devoid of points, with little effect on the IMSPE. This fact contradicts the oft-stated opinion that it is important not to have gaps in a practical design. However, Fig. 6 also shows that it is important that the direction between the twins be maintained as the ordinate, and not the abscissa, but, then again, the distance of separation, once the wrong choice of angle between the twins has been made, has little effect on the IMSPE. Thus, we were led to the observation, in the case of the twins of Fig. 5, the relevant variable, in practice, is the angle the twins make with the axes, and not the radial separation of the points from their midpoint. A hue plot showing the effect of moving one of the twins, while holding the twins' midpoint, and all other points fixed, is given in Fig. 7, below.

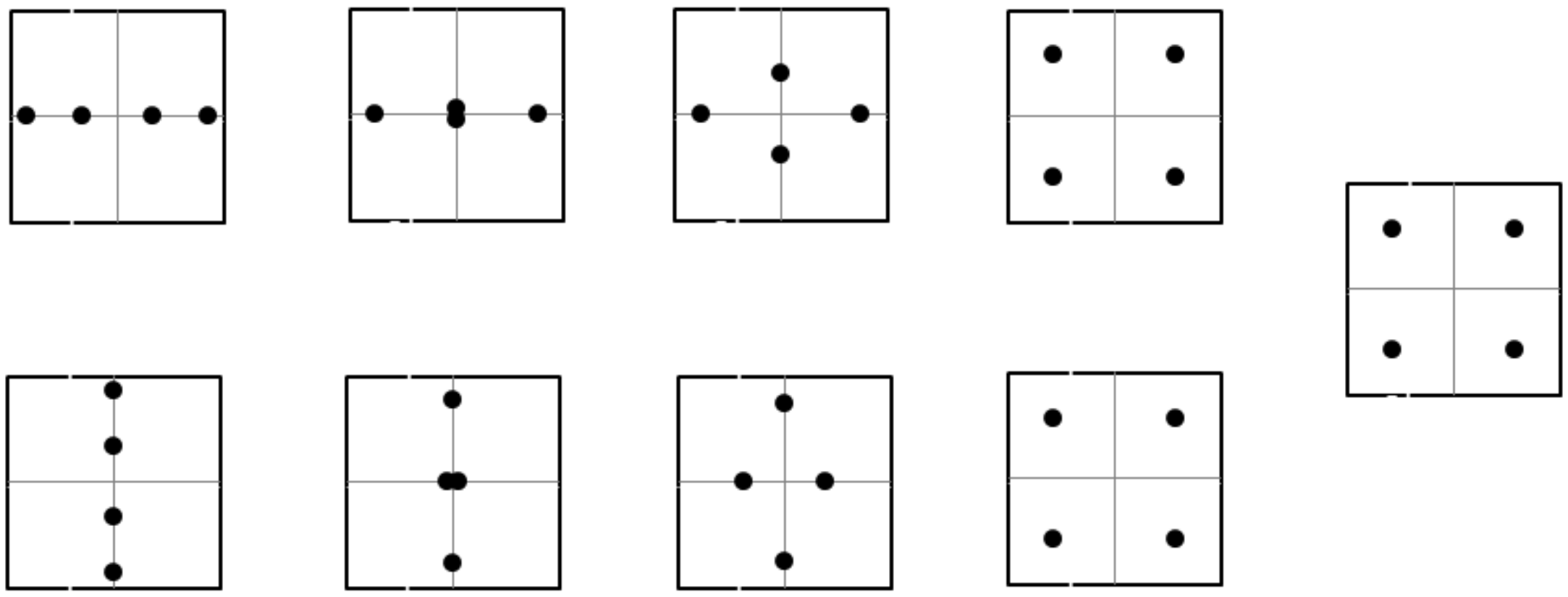

Fig. 4. Representative inversion-symmetric dot diagrams for the phases shown in Fig. 3 are shown, all with abscissas $x_1$ and ordinates $x_2$ on the domain $[-1,1]^2$. Points not noted as twins are non-proximal. Twin separations are exaggerated for visibility. From the upper-left corner and proceeding clockwise, the phases are the following: 4-in-lines along the $x_1$ axis; rhomboids, with twins on the $x_2$ axis; rhomboids, with two points on each of the axes; rectangles, with longer sides parallel with the $x_1$ axis; axis-oriented squares; and the same phases, in reverse order, with $x_1$ and $x_2$ interchanged.

| $x_1$ | $x_2$ |
|---|---|
| 0.000000… | 0.000000+δ |
| 0.000000… | 0.000000-δ |
| -0.767117… | 0.000000… |
| 0.767117… | 0.000000… |

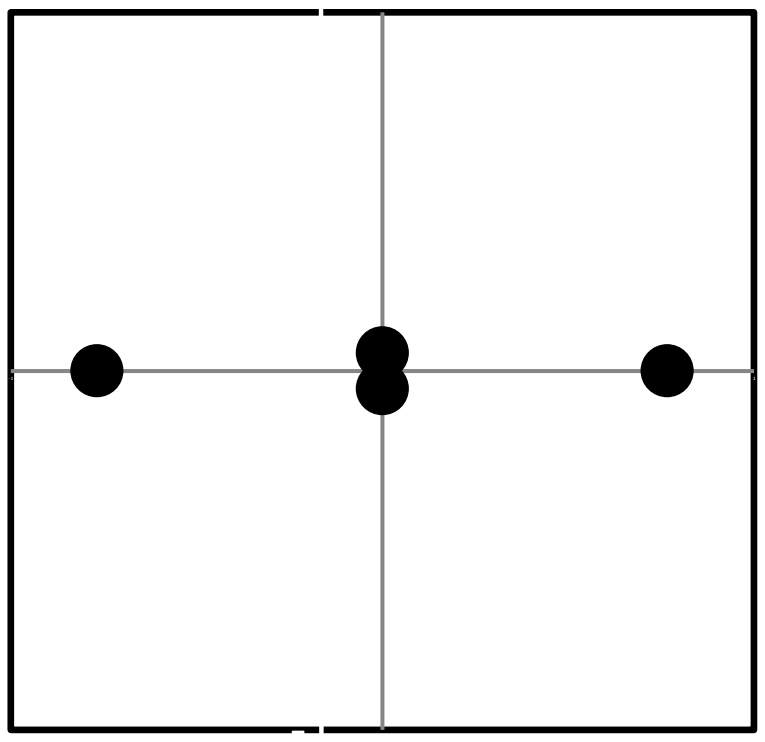

Fig. 5. The design listing (left), and the dot diagram (right), with exaggerated separation of the twins, for the N=4, two-factor, putatively IMSPE-optimal design problem mentioned in the text is shown.

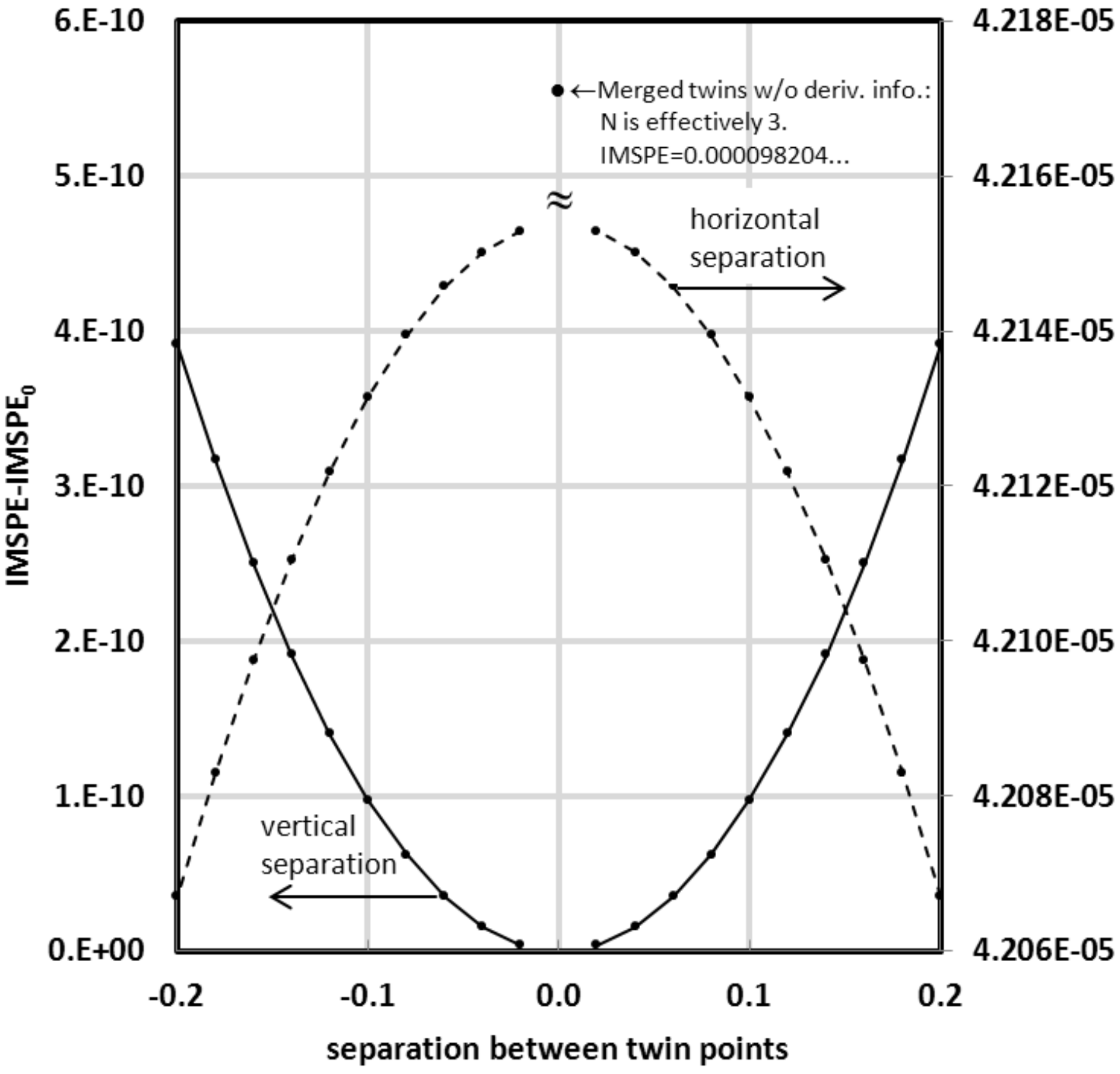

Fig. 6. This double-line plot shows the parabolic variation of the IMSPE of the design of Fig. 5, as a function of twin separation, with the left (resp., right) axis used for the case when the twins are separated along the $x_2$ (resp., $x_1$) axis. Dots represent computed values.

A tornado plot, Fig. 8, below, shows, for each of a large, representative random sample of designs, the $\log_{10}$-difference between each random design's IMSPE and the IMSPE=$IMSPE_0$ of the design of Fig. 5, plotted versus the maximum of the $\binom{4}{2} = 6$ half-$x_1$-distances between any two of the design's four design points. The tornado's shape reflects the parabolic approach to $IMSPE_0$.

## 7. Summary

We developed new, transportable and freely available, high-precision software for evaluating and searching for IMSPE-optimal designs on the design domain $[-1,1]^2$.

We used this software to search over a wide range of covariance-parameters and constructed a phase diagram with sharp boundaries for N=4, putatively optimal designs. The phase diagram included two contiguous domains of designs with twin points.

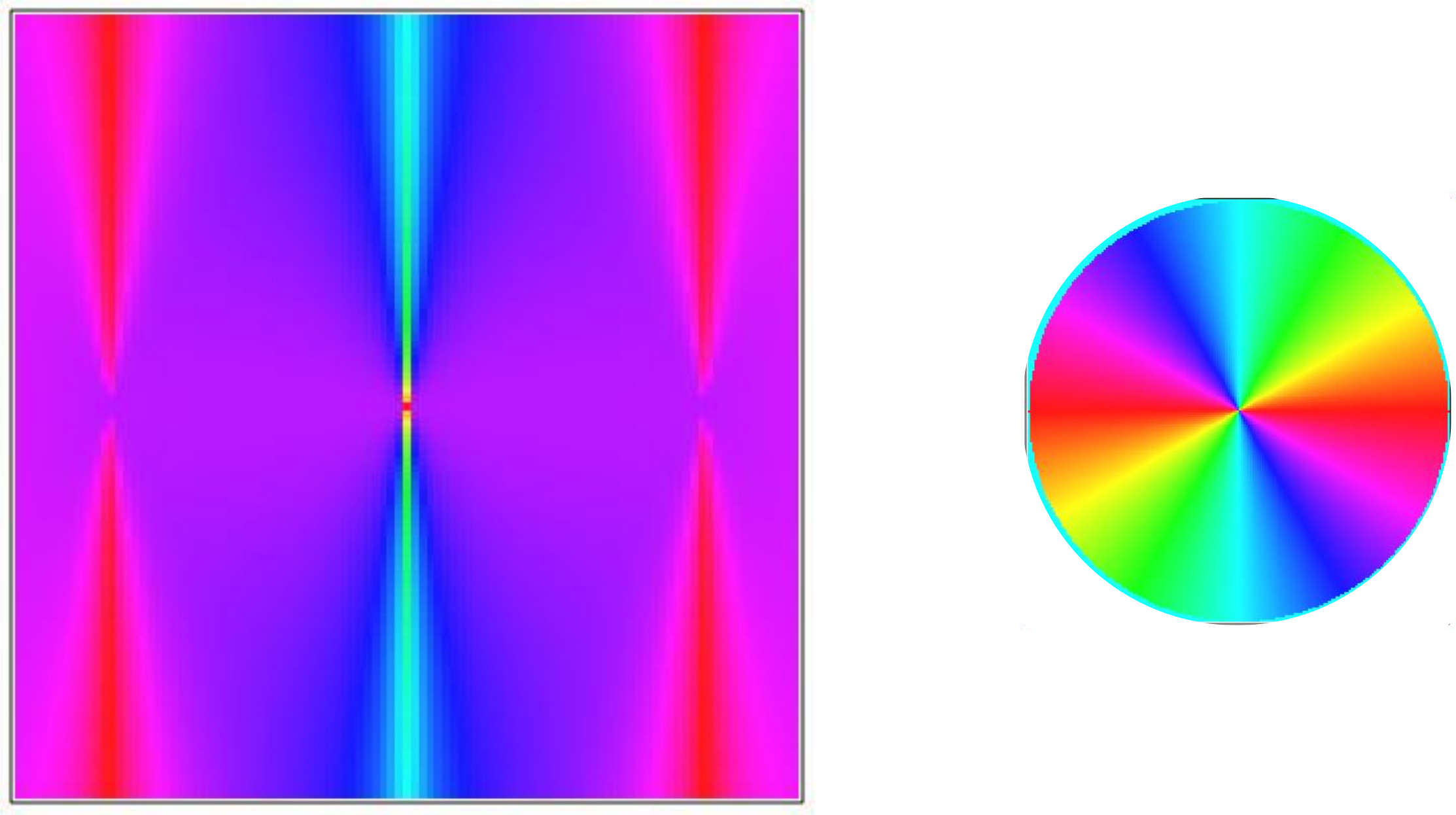

Fig. 7. The hue plot on the left shows $\log_{10}(\text{IMSPE-IMSPE}_0+10^{-16})$, coded as hue, over a $[-1,1]^2$ domain, with fixed design points at [±0.767117…,0], fixed twin-point midpoint at the origin, abscissa and ordinate representing the $x_1$ and $x_2$ coordinates of one of the twins, and the second twin forced to have inversion symmetry with respect to the first. The disk on the right represents the following color coding for the plot on the left: Considering the point where all colors in the disk meet as the origin, the increasing angle from a horizontal axis passing through the origin represents the $\log_{10}(\text{IMSPE-IMSPE}_0+10^{-16})$, from lowest value at zero radians to highest values at $\pi$ radians. Thus, the lowest $\log_{10}(\text{IMSPE-IMSPE0}+10^{-16})$ value is coded red, then the colors, as $\log_{10}(\text{IMSPE-IMSPE}_0+10^{-16})$ increases, pass in sequence through the traditional rainbow colors (red, orange, yellow, green, blue, and violet) and then continue cyclically with red. The plot on the left shows the following features: The minimum $\log_{10}(\text{IMSPE-IMSPE}_0+10^{-16})$ of the putatively optimal design is shown in red at the origin. The locally quadratic variation of IMSPE as the twins are separated vertically is shown by the rift valley running vertically from the origin. The local quadratic variation of IMSPE as the twins are separated horizontally is shown, with little color variation, moving horizontally across the plot. The red pixels near the corners of the square region represent high $\text{IMSPE-IMSPE}_0+10^{-16}$ values. The alert reader will recognize that the IMSPE should be multivalued at the origin of the plot on the left, but that the choice was made to use the lowest $\text{IMSPE-IMSPE}_0+10^{-16}$ value, there, and the origin was colored red.

We had the following evidence for the IMSPE-optimality of the Fig.-5 design: (1) Evaluation of more than one-million randomly generated designs, with the same covariance parameters, did not reveal any design with a lower IMSPE value. (2) All competitive designs fell on a tornado plot, consistent with a single, low-lying minimum; and (3) The well-defined phase diagram of Fig. 3 revealed no evidence of a design with lower IMSPE. Still, this evidence, while strong, was not mathematically conclusive, and we welcome others to complete the evidentiary case or provide a proof.

We showed that the IMSPE, local to a pair of twin points with fixed midpoint, is parabolic in the inter-point distance, demonstrating there is no loss of predictive information as the second of a pair of twins comes closer to the first.

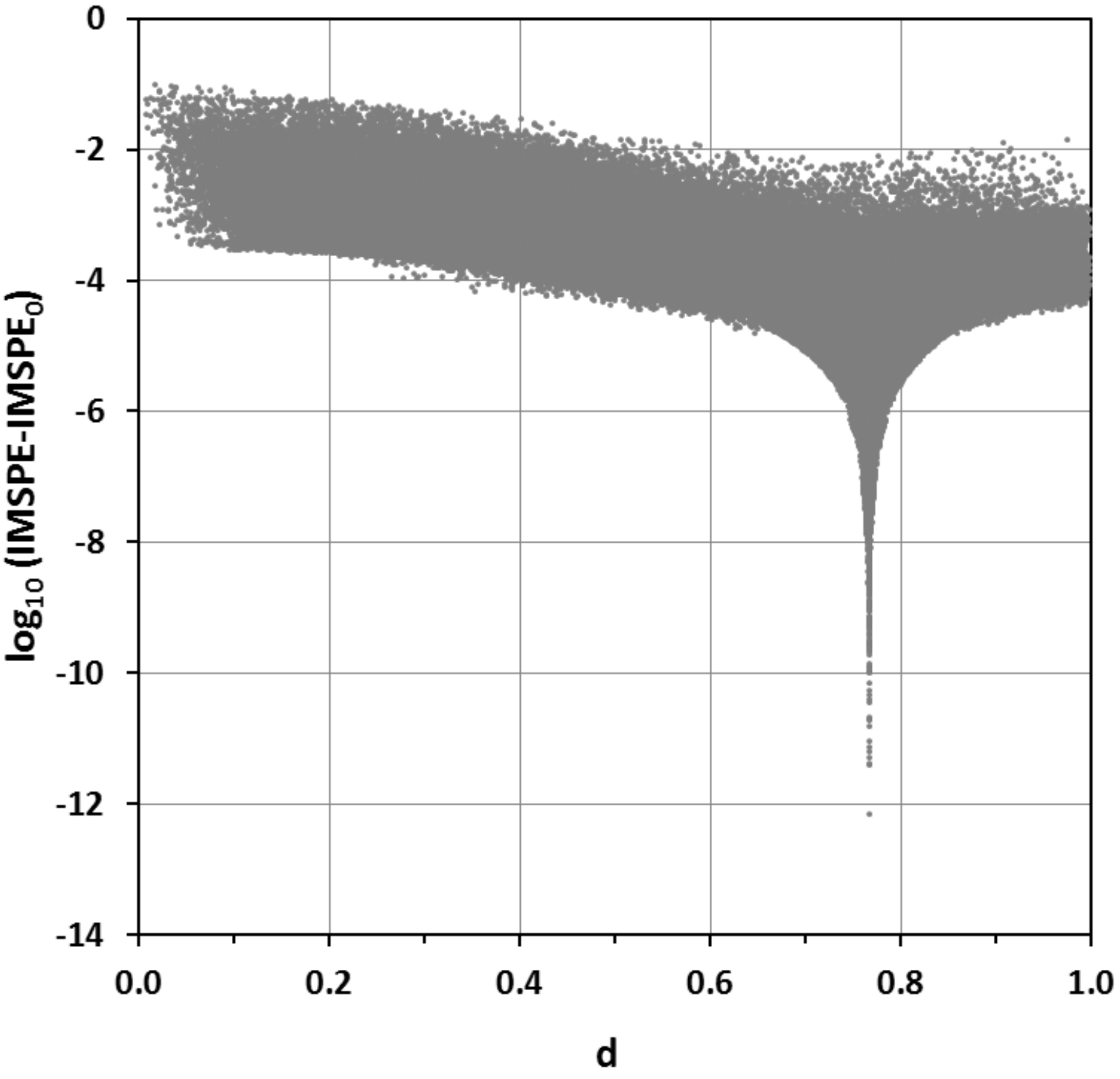

Fig. 8. This tornado plot shows, as grey dots, and for approximately one-million, randomly selected N=4 designs, and with the covariance parameters of the Fig.-4 design, the designs' $\log_{10}(\text{IMSPE-IMSPE}_0)$ vs. d values, where $\text{IMSPE}_0$ is the IMSPE of the Fig.-4 design, and d is the largest of the $\binom{4}{2}$ half-$x_1$-distances between any two of each design's four design points.

## 8. Conjectures

We make the following conjectures about exact, N-point designs over design domains $[-1,1]^D$, with finite Gaussian-covariance parameters, and N·D degrees of freedom (dof):

(1) There is no free-ranging, IMSPE-optimal design for D=1.
(2) The design in Fig. 5 is
  (a) IMSPE-optimal,
  (b) the IMSPE-optimal design with the smallest number of dof, and
  (c) the only IMSPE-optimal design with eight dof.
(3) There are no missing phases in the phase diagram in Fig. 3.

(4) In the log-log phase diagram of Fig. 3, the width of the rectangle-phase regions decreases as IMSPE is decreased, i.e. there is a bird's-beak shape to the phases with rectangles.

## 9. Concluding comments

Our hope is that others will use or extend our high-precision software, or develop their own, to carry out research on optimal design of computer experiments, unburdened by the following three prevalent, but mistaken, notions: (a) that all optimal-design points must be well spaced; (b) that designs with relatively large gaps in their dot diagrams, because of the gaps alone, are undesirable; and (c) that repulsion between design points [5], precludes ε-clustered designs.

We now cast aside our feigned disregard for statistical practice and assert it is possible to generate fast software for finding near-IMSPE-optimal designs, based on the established concepts of spread and coverage, augmented by the concept of irrelevance introduced in Section 6, and using well-known symmetry and tensor-product properties, along lines summarized recently by Plumlee [18].

Looking to the future, we mention that the IMSPE is an example of a possibly new class of real-valued functions of real variables that are $C^\infty$ everywhere, except for a finite number of essential discontinuities at the loci of the point clusters, with the various numbers and arrangements of clusters forming a nested hierarchy of homotopy classes. This class of functions is a subset of a larger class of low-degree-truncated rational functions. Research on these concepts is clearly indicated, including on what role they may play in the physical sciences.

We conclude with an indulgent, speculative paragraph. Our group has never gotten past SSW's Eq. 2.9, to any significant degree. Perhaps we can be forgiven our ideeé fixe on clusters, as there is "so much going on" when clusters and their homotopy-class interpretations arise from this statistical objective function, depending upon what information is carried by the simulation. This may be akin to what is "going on" in solutions of Schrödinger's equation, depending upon what information is accessible to an observer, in that latter equation's statistical interpretation.

*Hey, Vern, I heard you won the raffle for that free weekend in Portland. How was the big city?*
*Well, there was so much going on at the station, I never did make it to the town.*
*-- Down-East humor*

## 10. Revision history

V2: Ref. 18 was added to the References; a comment on the potentially new class of functions was added to the penultimate paragraph of the ultimate section, where the mention of Panlevé transcendents was removed; and minor typographical changes were made.
V3: In Fig. 3, the boundaries of both rhombus-with-twins phases were corrected; in the Appendix, all $\sigma_Z^2$ were changed to unity; and Ref. 1 was corrected. In Sec. 6, the sentence mentioning "topological" was deleted.
V4: All instances of "barycenter" were changed to "midpoint;" one errant instance of "$\sigma_Z^4$" was removed; and blocks of matrix $\boldsymbol{R}$ in the Appendix were defined, and their integrals written out correctly.

## Acknowledgments

We thank Profs. Hermann Schichl and Arnold Neumaier of the Univ. of Vienna's Mathematics Faculty for a discussion of homotopy theory; Prof. Erin Leatherman of West Virginia University for enlightening conversations; Keith Hall (ret.) of Knolls Atomic Power Laboratory, Niskayuna, NY for his assistance in computing the dynamics of small, non-linear systems; and Prof. George Ruppeiner of New College of Florida for providing an example of a low-degree-truncated rational function arising in physics.

## References

1. Jerome Sacks, Susannah B. Schiller, and William J. Welch, "Design for Computer Experiments," *Technometrics* **31** (1), pp. 41-47 (1989).

2. Ben Haaland and Peter Z. G. Qian, "Accurate Emulators for Large-Scale Computer Experiments," *Ann. Statist.* **39** (6), pp. 2974-3002 (2011), with special attention to pp. 2981-2982.

3. Gemma Stephenson (2010) "Using Derivative Information in the Statistical Analysis of Computer Models," University of Southampton, School of Ocean and Earth Science, Ph.D. Dissertation, pp. i-197. Specific attention is drawn to the following pages and paragraphs: p. 88, ¶ 3; p. 143; p. 145, ¶ 1; and p. 147, ¶ 2 through the end of that page.

4. Erin Rae Leatherman (2013) "Optimal Predictive Designs for Experiments that Involve Computer Simulators," Ohio State University, Ph.D. Dissertation, pp. i-396.

5. Luc Pronzato and Werner G. Müller, "Design of Computer Experiments: Space Filling and Beyond," *Stat. Comput.* **22**, pp. 681-701 (2012).

6. Selden B. Crary, "Statistical Design and Analysis of Computer Experiments for the Generation of Parsimonious Metamodels," Published in *Design, Test, Integration, and Packaging of MEMS/MOEMS 2001*, B. Courtois, J. M. Karam, S. P. Levitan, K. W. Markus, A. A. O. Tay, and J. A. Walker, Eds., Proceeds. SPIE **4408**, pp. 29-39 (2001).*

7. Selden B. Crary, David M. Woodcock, and Andreas Hieke, "Designing Efficient Computer Experiments for Metamodel Generation," Published in the *Proceedings of the Fourth International Conference on Modeling of Microsystems, MSM 2001*, Hilton Head, SC, March 19-21, 2001, pp. 132-135.*

8. Selden B. Crary, "Design of Computer Experiments for Metamodel Generation," *Special Invited Issue of Analog Integrated Circuits and Signal Processing* **32**, pp. 7-16 (2002).*

9. Selden B. Crary and Rachel Johnson, "Validation of the Twin-Point-Design Concept in the Design of Computer Experiments," Section on Statistical Computing – JSM 2011, pp. 5495-5505.*

10. Jerome Sacks, William J. Welch, Toby J. Mitchell, and Henry P. Wynn, "Design and Analysis of Computer Experiments," *Statistical Science* **4**, pp. 409-423 (1989).

11. Werner Müller and Dale L. Zimmerman, "Optimal Designs for Variogram Estimation," *Environmetics* **10**, pp. 23-37 (1999).

12. Zhengyuan Zhu and Hao Zhang, "Spatial Sampling Design Under the Infill Asymptotic Framework," *Environmetics* **17**, pp. 323-337 (2006).

13. Dale L. Zimmerman, "Optimal Network Design for Spatial Prediction, Covariance Parameter Estimation, and Empirical Prediction," *Environmetics* **17** (6), pp. 635-652 (2006).

14. Kathryn M. Irvine, Alix I. Gittelman, and Jennifer A. Hoeting, “Spatial Designs and Properties of Spatial Correlation: Effects on Covariance Estimation,” *J. Agric. Biol. Environ. Stat.* **12** (4), pp. 450-469 (2007).

15. Zhengyuan Zhu and Michael L. Stein, “Spatial Sampling Design for Parameter Estimation of the Covariance Function,” *J. Statist. Plann. Inference* **134**, (2), pp. 583–603 (2005).

16. Selden B. Crary, “New Research Directions in Computer Experiments: ε-Clustered Designs,” Spring Research Conference (SRC 2012), Cambridge, MA, June 13-15, 2012, presentation published in *JSM Proceedings*, Statistical Computing Section, Alexandria, VA, USA: ASA, pp. 5692-5706 (2012). URL: https://www.amstat.org/sections/srms/proceedings/y2012/files/400245_500701.pdf.*

17. Fred J. Hickernell, private communication, mentioned with permission.

18. Matthew Plumlee, “Fast Prediction of Deterministic Functions Using Sparse Grid Experimental Designs,” *J. Amer. Statist. Assoc*. **109** (508), pp. 1581-1591 (2014).

*Revisions are available from the first author at email address: selden_crary (at) yahoo (dot) com.

## Appendix. Algebra for formulas used in *evalIMSPE2D* and *minIMSPEccd2D*

**Definitions:** The foundational paper by Sacks et al. [1] provides the theoretical background and notation. “IMSE,” which was used in [1] is a synonym for “IMSPE” (integrated mean-squared prediction error), which was used in [4].

*N*-point, *D*-factor design:
$\{\boldsymbol{x_1}, \boldsymbol{x_2}, \cdots, \boldsymbol{x_N}\} = \{(x_{1,1}, x_{1,2}, \cdots, x_{1,D}),\ (x_{2,1}, x_{2,2}, \cdots, x_{2,D}), \cdots, (x_{N,1}, x_{N,2}, \cdots, x_{N,D})\} \quad -1 \le x_{m,n} \le 1.$

We start from Eq. 2.9 of [1]. $IMSE/\sigma_z^2 = 1 - tr(\boldsymbol{L}^{-1}\boldsymbol{R})$, where $(N+1)x(N+1)$, dimensionally homogeneous, symmetric matrices $\boldsymbol{L}$ (from “left,” note well, not “lower,” matrix and $\boldsymbol{R}$ (from “right” matrix in the same equation) are defined as

$$\boldsymbol{L} \equiv \begin{pmatrix} 0 & | & 1 & \cdots & 1 \\ - & | & - & - & - \\ 1 & | & & & \\ \vdots & | & & \boldsymbol{V} & \\ 1 & | & & & \end{pmatrix} \text{ and } \boldsymbol{R} \equiv \frac{1}{2^D}\int_{-1}^{1}\int_{-1}^{1}\cdots\int_{-1}^{1} \begin{pmatrix} 1 & | & v_1 & v_2 & \cdots & v_N \\ -- & | & -- & -- & -- & -- \\ v_1 & | & v_1^2 & v_1v_2 & \cdots & v_1v_N \\ v_2 & | & v_1v_2 & v_2^2 & \cdots & v_2v_N \\ \vdots & | & \vdots & \vdots & \ddots & \vdots \\ v_N & | & v_1v_N & v_2v_N & \cdots & v_N^2 \end{pmatrix} dx_1 dx_2 \cdots dx_D,$$

where $\boldsymbol{V}$ is the correlation matrix, $V_{i,j} = exp\left[-\sum_{k=1}^{D}\theta_k\left(x_{i,k} - x_{j,k}\right)^2\right]$; $\boldsymbol{v}_i = exp\left[-\sum_{k=1}^{D}\theta_k\left(x_{i,k} - x_k\right)^2\right]$; where $(\boldsymbol{x}_i, \boldsymbol{x}_j)$ $i,j = 1 \cdots N$ are the design points; $\boldsymbol{\theta}$ is a $Dx1$ vector of non-negative covariance parameters; and the indices $i$ and $j$ of the one pair of twin points are $N-1$ and $N$.

The integrals in $\boldsymbol{R}$ are based on the following two integrals:

$$I_{1;a} \equiv \frac{1}{2}\int_{-1}^{1} exp[-\theta(a-x)^2]\,dx = \sqrt{\frac{\pi}{16\theta}}\begin{Bmatrix} erf\left[\sqrt{\theta}(1+a)\right] \\ +erf\left[\sqrt{\theta}(1-a)\right] \end{Bmatrix}, \text{ and}$$

$$I_{2;a,b} \equiv \frac{1}{2}\int_{-1}^{1} exp\{-\theta[(a-x)^2 + (b-x)^2]\}\,dx = \sqrt{\frac{\pi}{32\theta}}\begin{Bmatrix} erf\left[\sqrt{2\theta}\left(1+\frac{a+b}{2}\right)\right] \\ +erf\left[\sqrt{2\theta}\left(1-\frac{a+b}{2}\right)\right] \end{Bmatrix} exp\left[-\frac{\theta(a-b)^2}{2}\right],$$

where *a* and *b* are arbitrary real constants.

**Integrals appearing in *R*:**

$$\tilde{I}_{1;i} \equiv \frac{1}{2^D}\int_{-1}^{1}\int_{-1}^{1}\cdots\int_{-1}^{1} v_i \, dx_1 dx_2 \cdots dx_D = \frac{1}{2^D}\int_{-1}^{1}\int_{-1}^{1}\cdots\int_{-1}^{1} exp\left[-\sum_{k=1}^{D}\theta_k \left(x_{i,k}-x_k\right)^2\right] dx_1 dx_2 \cdots dx_D$$

$$= \prod_{k=1}^{D}\sqrt{\frac{\pi}{16\theta_k}}\left\{\begin{matrix} erf\left[\sqrt{\theta_k}\left(1+x_{i,k}\right)\right] \\ +erf\left[\sqrt{\theta_k}\left(1-x_{i,k}\right)\right]\end{matrix}\right\}, \text{ and}$$

$$\tilde{I}_{2;i,j} \equiv \frac{1}{2^D}\int_{-1}^{1}\int_{-1}^{1}\cdots\int_{-1}^{1} v_i \, v_j dx_1 dx_2 \cdots dx_D = \frac{1}{2^D}\int_{-1}^{1}\int_{-1}^{1}\cdots\int_{-1}^{1} exp\left[-\sum_{k=1}^{D}\theta_k\left[\left(x_{i,k}-x_k\right)^2+\left(x_{j,k}-x_k\right)^2\right]\right] dx_1 dx_2 \cdots dx_D$$

$$= \prod_{k=1}^{D}\sqrt{\frac{\pi}{32\theta_k}}\left\{\begin{matrix} erf\left[\sqrt{2\theta_k}\left(1+\frac{x_{i,k}+x_{j,k}}{2}\right)\right] \\ +erf\left[\sqrt{2\theta_k}\left(1-\frac{x_{i,k}+x_{j,k}}{2}\right)\right]\end{matrix}\right\} exp\left[-\frac{\theta_k\left(x_{i,k}-x_{j,k}\right)^2}{2}\right].$$

Symmetric matrix ***R*** can be expressed in terms of evaluations of the error function, via related functions $S_l(\boldsymbol{x_i},\boldsymbol{\theta}) \quad l = 1,2$, as follows:

$$\boldsymbol{R} = \begin{pmatrix} 1 & | & \cdots & S_1(\boldsymbol{x_j},\boldsymbol{\theta}) & \cdots \\ -- & | & -- & ------------------ & -- \\ \vdots & | & \cdots & \vdots & \cdots \\ \cdot & | & \vdots & S_2\left(\frac{\boldsymbol{x_i}+\boldsymbol{x_j}}{2},\boldsymbol{\theta}\right) exp - \frac{\sum_{k=1}^{D}\theta_k\left(x_{i,k}-x_{j,k}\right)^2}{2} & \vdots \\ \vdots & | & \cdots & \vdots & \cdots \end{pmatrix},$$

where the row and column indices of ***R*** run from *0* through *N*, the first indices of the design run from *1* through *N*, and

$$S_\ell(\boldsymbol{x_i},\boldsymbol{\theta}) = \prod_{k=1}^{D}\sqrt{\frac{\pi}{16\ell\theta_k}}\left\{\begin{matrix} erf\left[\sqrt{\ell\theta_k}\left(1+x_{i,k}\right)\right] \\ +erf\left[\sqrt{\ell\theta_k}\left(1-x_{i,k}\right)\right]\end{matrix}\right\}, \; \ell = 1,2.$$

In particular, for $D = 2$,

$$\boldsymbol{R} = \begin{bmatrix} 1 & | & \cdots & \left(\frac{\pi}{16}\right)\sqrt{\frac{1}{\theta_1\theta_2}}\begin{pmatrix} \begin{Bmatrix} erf\left[\sqrt{\theta_1}\left(1+x_{j,1}\right)\right] \\ +erf\left[\sqrt{\theta_1}\left(1-x_{j,1}\right)\right] \end{Bmatrix} \\ \cdot\begin{Bmatrix} erf\left[\sqrt{\theta_2}\left(1+x_{j,2}\right)\right] \\ +erf\left[\sqrt{\theta_2}\left(1-x_{j,2}\right)\right] \end{Bmatrix} \end{pmatrix} & \cdots \\ -- & | & -- & ------------------------------ & -- \\ \cdot & | & \cdots & \vdots & \cdots \\ \cdot & | & \cdots & \left(\frac{\pi}{32}\right)\sqrt{\frac{1}{\theta_1\theta_2}}\begin{pmatrix} \begin{Bmatrix} erf\left[\sqrt{2\theta_1}\left(1+\frac{x_{i,1}+x_{j,1}}{2}\right)\right] \\ +erf\left[\sqrt{2\theta_1}\left(1-\frac{x_{i,1}+x_{j,1}}{2}\right)\right] \end{Bmatrix} \\ \cdot\begin{Bmatrix} erf\left[\sqrt{2\theta_2}\left(1+\frac{x_{i,2}+x_{j,2}}{2}\right)\right] \\ +erf\left[\sqrt{2\theta_2}\left(1-\frac{x_{i,2}+x_{j,2}}{2}\right)\right] \end{Bmatrix} \end{pmatrix} exp\left\{-\frac{\begin{bmatrix} \theta_1\left(x_{i,1}-x_{j,1}\right)^2 \\ +\theta_2\left(x_{i,2}-x_{j,2}\right)^2 \end{bmatrix}}{2}\right\} & \cdots \\ \cdot & | & \cdot & \vdots & \cdots \end{bmatrix}. \qquad (A.1)$$

For $D = 2$ cases with exactly one twin point, and naming the blocks of $\boldsymbol{R}$ as

$$\boldsymbol{R} = \begin{pmatrix} \boldsymbol{A} & | & \boldsymbol{B} & | & \boldsymbol{C} \\ -- & | & -- & | & -- \\ \cdot & | & \boldsymbol{D} & | & \boldsymbol{E} \\ -- & | & -- & | & -- \\ \cdot & | & \cdot & | & \boldsymbol{F} \end{pmatrix}.$$

Eq. A.1 can be written as the following symmetric matrix, after permuting indices so the twins have the two highest indices, viz., $N - 1$ and $N$, and after defining the midpoint of the twins as $\boldsymbol{x_t} \equiv \left(x_{m,1}, x_{m,2}\right)$:

$$
\boldsymbol{R} = \left(\begin{array}{cc|ccc|cc}
1 & | & \cdots & R_{0,j} & \cdots & R_{0,N-1} & R_{0,N} \\
\hline
\cdot & | & R_{1,1} & \cdots & R_{1,N-2} & R_{1,N-1} & R_{1,N} \\
\cdot & | & \vdots & \ddots & \vdots & \vdots & \vdots \\
\cdot & | & \cdot & \cdots & R_{N-2,N-2} & R_{N-2,N-1} & R_{N-2,N} \\
\hline
\cdot & | & \cdot & \cdots & \cdot & R_{N-1,N-1} & R_{N-1,N} \\
\cdot & | & \cdot & \cdots & \cdot & \cdot & R_{N,N}
\end{array}\right).
$$

| Block names | Integrals | |
| --- | --- | --- |
| ***B*** | $R_{0,j}$ | $= \left(\frac{\pi}{16}\right)\sqrt{\frac{1}{\theta_1\theta_2}}\begin{Bmatrix} erf\left[\sqrt{\theta_1}\left(1+x_{j,1}\right)\right] \\ +erf\left[\sqrt{\theta_1}\left(1-x_{j,1}\right)\right]\end{Bmatrix}\begin{Bmatrix} erf\left[\sqrt{\theta_2}\left(1+x_{j,2}\right)\right] \\ +erf\left[\sqrt{\theta_2}\left(1-x_{j,2}\right)\right]\end{Bmatrix}.$ |
| ***C*** | $R_{0,\substack{N-1\\N}}$ | $= \left(\frac{\pi}{16}\right)\sqrt{\frac{1}{\theta_1\theta_2}}\begin{Bmatrix} erf\left[\sqrt{\theta_1}\left(1+x_{m,1}\pm\delta_1\right)\right] \\ +erf\left[\sqrt{\theta_1}\left(1-x_{m,1}\mp\delta_1\right)\right]\end{Bmatrix}\begin{Bmatrix} erf\left[\sqrt{\theta_2}\left(1+x_{m,2}\pm\delta_2\right)\right] \\ +erf\left[\sqrt{\theta_2}\left(1-x_{m,2}\mp\delta_2\right)\right]\end{Bmatrix}.$ |
| ***D*** | $R_{i,j}$ | $= \left(\frac{\pi}{32}\right)\sqrt{\frac{1}{\theta_1\theta_2}}\begin{Bmatrix} erf\left[\sqrt{\theta_1}\left(1+x_{i,1}\right)\right] \\ +erf\left[\sqrt{\theta_1}\left(1-x_{i,1}\right)\right]\end{Bmatrix}\begin{Bmatrix} erf\left[\sqrt{\theta_2}\left(1+x_{i,2}\right)\right] \\ +erf\left[\sqrt{\theta_2}\left(1-x_{i,2}\right)\right]\end{Bmatrix}.$ |
| ***E*** | $R_{i,\substack{N-1\\N}}$ | $= \left(\frac{\pi}{32}\right)\sqrt{\frac{1}{\theta_1\theta_2}}\begin{Bmatrix} erf\left[\sqrt{2\theta_1}\left(1+\frac{x_{i,1}+x_{m,1}\pm\delta_1}{2}\right)\right] \\ +erf\left[\sqrt{2\theta_1}\left(1-\frac{x_{i,1}+x_{m,1}\pm\delta_1}{2}\right)\right]\end{Bmatrix}\begin{Bmatrix} erf\left[\sqrt{2\theta_2}\left(1+\frac{x_{i,2}+x_{m,2}\pm\delta_2}{2}\right)\right] \\ +erf\left[\sqrt{2\theta_2}\left(1-\frac{x_{i,2}+x_{m,2}\pm\delta_2}{2}\right)\right]\end{Bmatrix} exp\left\{-\frac{\begin{bmatrix}\theta_1\left(x_{i,1}-x_{m,1}\mp\delta_1\right)^2 \\ +\theta_2\left(x_{i,2}-x_{m,2}\mp\delta_2\right)^2\end{bmatrix}}{2}\right\}.$ |
| ***F*** | $R_{\substack{N-1,N-1\\N,N}}$ | $= \left(\frac{\pi}{32}\right)\sqrt{\frac{1}{\theta_1\theta_2}}\begin{Bmatrix} erf\left[\sqrt{2\theta_1}\left(1+x_{m,1}\pm\delta_1\right)\right] \\ +erf\left[\sqrt{2\theta_1}\left(1-x_{m,1}\mp\delta_1\right)\right]\end{Bmatrix}\begin{Bmatrix} erf\left[\sqrt{2\theta_2}\left(1+x_{m,2}\pm\delta_2\right)\right] \\ +erf\left[\sqrt{2\theta_2}\left(1-x_{m,2}\mp\delta_2\right)\right]\end{Bmatrix}.$ |
| ***F*** | $R_{N-1,N}$ | $= \left(\frac{\pi}{32}\right)\sqrt{\frac{1}{\theta_1\theta_2}}\begin{Bmatrix} erf\left[\sqrt{2\theta_1}\left(1+x_{t,1}\right)\right] \\ +erf\left[\sqrt{2\theta_1}\left(1-x_{t,1}\right)\right]\end{Bmatrix}\begin{Bmatrix} erf\left[\sqrt{2\theta_2}\left(1+x_{m,2}\right)\right] \\ +erf\left[\sqrt{2\theta_2}\left(1-x_{m,2}\right)\right]\end{Bmatrix} exp(-2\boldsymbol{\theta}\cdot\boldsymbol{\delta}^2),$ |

where $\boldsymbol{\theta}\cdot\boldsymbol{\delta}^2$ in $R_{N-1,N}$ is the dot product of vectors $\boldsymbol{\theta} \equiv (\theta_1, \theta_2)$ and $\boldsymbol{\delta}^2 \equiv (\delta_1^2, \delta_2^2)$, and $\boldsymbol{\delta}$ is the vector from the twins' midpoint to the first of the twins, viz. the point with index $N-1$. It may be possible to save precision by using $\boldsymbol{\delta}^2$ directly, instead of computing it via subtractions of moderately sized numerical values.